\documentclass{agujournal2019}
\usepackage{url} 
\usepackage{lineno}
\usepackage[inline]{trackchanges} 
\usepackage{soul}
\usepackage{amsmath}
\usepackage{amsfonts}
\usepackage{amssymb}
\usepackage{bm}
\usepackage{braket}
\nolinenumbers
\renewcommand{\linenumbers}{}  

\renewcommand{\vec}[1]{\bm{#1}} 

\draftfalse

\journalname{JGR: Solid Earth}

\begin{document}

\title{Ab initio calculated diamagnetic and paramagnetic susceptibility of carbonate minerals}

\authors{Jaroslav Hamrle\affil{1,2}, Matěj Machek\affil{3}, Vladimír K.~Kusbach\affil{3}, Zuzana Roxerová\affil{3}}

\affiliation{1}{Faculty of Nuclear Sciences and Physical Engineering, Czech Technical University, Trojanova 13, Prague 12000, Czech Republic}
\affiliation{2}{Faculty of Mathematics and Physics, Charles University, Ke Karlovu 5, Prague 12116, Czech Republic}
\affiliation{3}{ Institute of Geophysics of the Czech Academy of Sciences, Boční II/1401, Prague 14100, Czech Republic}

\correspondingauthor{Jaroslav Hamrle}{jaroslav.hamrle@fjfi.cvut.cz}



\begin{keypoints}
\item We calculate the diamagnetic susceptibility of carbonate minerals ab initio
\item We calculate the paramagnetic susceptibility of Fe-doped carbonate minerals ab initio
\item We quantify the contribution of spin, spin-orbit and orbital magnetisation to magnetic susceptibility and its anisotropy
\end{keypoints}

%
%

%
%


\begin{abstract}
The anisotropy of magnetic susceptibility (AMS) is a widely used tool to infer rock fabrics, yet quantitative interpretation is limited by sparse single-crystal magnetic properties for rock-forming minerals and by the difficulty of separating intrinsic diamagnetism from impurity-related magnetism. Here we use density-functional theory (DFT) combined with perturbation theory to compute the diamagnetic susceptibility and AMS of calcite-group carbonates (calcite, magnesite, and dolomite) and to quantify the additional paramagnetic contribution from transition-metal doping.
For pure calcite, the calculated susceptibility and its anisotropy are in good agreement with published single-crystal measurements, validating the ab initio approach for diamagnetic phases. We provide improved intrinsic diamagnetic reference values for magnesite and dolomite, for which experimental susceptibilities are commonly affected by magnetic impurities. To address impurity effects explicitly, we model Fe and Mn substitution in calcite using supercells. The computed spin moments reproduce expected high-spin states (Fe$^{2+}$, $S=2$; Mn$^{2+}$, $S=5/2$) and the measured increase of susceptibility with Fe and Mn content. A preliminary calculation of the orbital response gives a susceptibility anisotropy per Fe atom comparable to the experimental value, tied to the crystallographic $c$-axis. Its strong temperature dependence indicates single-ion anisotropy of Fe$^{2+}$, highlighting orbital magnetisation as a key but numerically challenging ingredient for modelling of paramagnetic AMS in carbonates.
\end{abstract}

\section*{Plain Language Summary}
Magnetic susceptibility describes how a material responds to a weak magnetic field. In many rocks, this response is slightly different in different directions (anisotropy of magnetic susceptibility, AMS), and geoscientists use AMS to infer how minerals are oriented and how rocks formed or deformed. For carbonate minerals such as calcite, magnesite, and dolomite, AMS interpretation is often uncertain because the intrinsic (diamagnetic) signal is small and can be overprinted by tiny amounts of paramagnetic impurities.
In this study we use quantum-mechanical calculations to compute the diamagnetic susceptibility and AMS of calcite-group carbonates, and we also model the additional paramagnetic contribution caused by Fe and Mn substitutions. For pure calcite, the calculated susceptibility and AMS match single-crystal measurements, showing that the diamagnetic AMS can be predicted reliably. We then provide intrinsic reference values for magnesite and dolomite, which are difficult to obtain from natural samples. Finally, our calculations reproduce the magnetic moments of Fe and Mn dopants and the measured increase of susceptibility with Fe and Mn content. Preliminary calculations link the strong AMS increase with Fe content to the orbital motion of electrons of individual Fe ions, oriented by the crystal c-axis, rather than to magnetic interaction between dopant atoms.

%
%

\section{Introduction}

The anisotropy of magnetic susceptibility (AMS) is a widely applied method for characterising rock fabrics, e.g.~\cite{Hrouda1982}; \cite{Tarling1993}; \cite{Borradaile2004}; \cite{Borradaile2010}; \cite{Morris2019}; \cite{Fodor2020}; \cite{Narloch2021} and fabric development in analogue models \cite{Kratinova2010}; \cite{Zavada2009}; \cite{Martin2023}; \cite{Roxerova2023}. The AMS reflects an integrated magnetic signal produced by crystallographic and shape preferred orientation of all mineral grains forming a rock microstructure \cite{Borradaile1981}. Rock microstructures often exhibit considerable complexity, resulting from the interplay of distinct microstructural features and the superposition of subfabrics with varying orientations and intensities. These features reflect a range of processes: sedimentation, diagenesis, magma flow, deformation and metamorphism \cite{Hirt2012}; \cite{Kusbach2019}; \cite{Machek2019}; \cite{Pares2004}; \cite{Haerinck2013}.

Numerical modelling of AMS based on microstructural data has emerged as a valuable tool for interpreting geological processes responsible for AMS development and identifying the superposition of subfabrics \cite{Arbaret2000}; \cite{Kusbach2019}; \cite{Biedermann2020Challenges}, including carbonate rocks (e.g.~\cite{deWall2000}; \cite{Schmidt2009}). However, significant challenges arise from limitations in microstructural analysis and incomplete knowledge of mineral magnetic properties \cite{Biedermann2018}; \cite{Kusbach2019}; \cite{Biedermann2020FinIrrSDA}. The existing dataset of experimentally single-crystal magnetic properties of rock-forming minerals remains sparse, particularly for mineral species of variable chemical composition \cite{Biedermann2018}. Experimental AMS estimation in a single crystal of rock-forming minerals has been performed by using both low-field and high-field measurement techniques \cite{Biedermann2018}. Low-field measurements are often affected by the presence of inclusions and exsolutions \cite{Lagroix2000}, while high-field torque methods, though less influenced by these features, are constrained by limited access to the specialised instruments. Sensitivity decreases with a smaller sample size in both methods, whereas larger samples are more likely to contain inclusions and exsolution features. Within the calcite group, magnetic anisotropy has been experimentally constrained with reasonable reliability only for calcite and siderite \cite{Schmidt2006}.
In principle, these limitations could be addressed by theoretical estimates of the magnetic anisotropy of minerals through quantum-mechanical calculations. Recent advancements in the ab initio computation methods, combined with the rapid growth in computational power, have led to significant advances in theoretical mineral physics. In the Earth sciences, ab-initio calculations are frequently employed for a variety of purposes, including constructing mineralogical models and estimating the physical properties of the deep Earth (for review see \cite{Tsuchiya2020}. These calculations are also used to study mineral-water or mineral-melt interactions (e.g. \cite{Midgley2021}; \cite{Ulian2021}; \cite{Guo2024}), the solubility and crystallization of critical metals from hydrothermal fluids (\cite{Sherman2010}, \cite{Brugger2016}, \cite{Mei2020}), incorporation of OH group and rare earth elements elements in crystal structures, and the fractionation of elements and isotopes in both deep and surface environments (e.g. \cite{Karki2018}; \cite{Liu2021}; \cite{Hoare2022}; \cite{Kim2022}; \cite{Hu2023}). Despite these advancements,  the magnetic anisotropy of mineral phases remains unexplored using this approach. For the calcite group, which is the focus of this study, most first-principles calculations have primarily investigated optical and elastic properties under high pressure  (\cite{Zhao2009}; \cite{Ayoub2011}; \cite{Bakri2011}; \cite{Brik2011}; \cite{Hossain2011}; \cite{Marcondes2016}; \cite{Fu2017}; \cite{Solomatova2018}; \cite{Zhuravlev2021}; \cite{Ulian2022}).
The implementation of ab-initio calculated diamagnetic susceptibility is based on perturbation theory. First, it calculates the current density induced by the external magnetic field. Knowing the current density, the induced magnetic field is calculated using the Biot-Savart law. The magnetic field can be calculated locally (e.g. on the nuclei' position, providing NMR screening), or its average value over the unit cell, providing diamagnetic susceptibility. This method was implemented in the Wien2k code, within the package NMR \cite{Laskowski2012}, yielding good agreement between experimental and ab initio-calculated magnetic susceptibilities of various salts \cite{Laskowski2014}. Alternative calculation methods of magnetic susceptibility were developed, based on geometrical curvature of Bloch electronic states, omitting calculation of current density \cite{Ogata2015}, \cite{Gao2015}. However, those approaches have not been numerically implemented in DFT codes yet.
To test the application of theoretical quantum-mechanical calculations for determining the magnetic anisotropy of diamagnetic mineral phases, we have selected the calcite group mineral series. This series features a relatively simple crystal structure where the exchange of Ca, Mg, Fe and Mn cations within the crystal structure induces variation in magnetic properties. The diamagnetic properties of calcite, dolomite and magnesite are calculated and the influence of Fe and Mn doping is tested. The results of theoretical calculations are compared with the existing experimental data.

\section{Properties of Calcite -- Magnesite mineral series}

\begin{figure}
    \centering
    \includegraphics[width=1\linewidth]{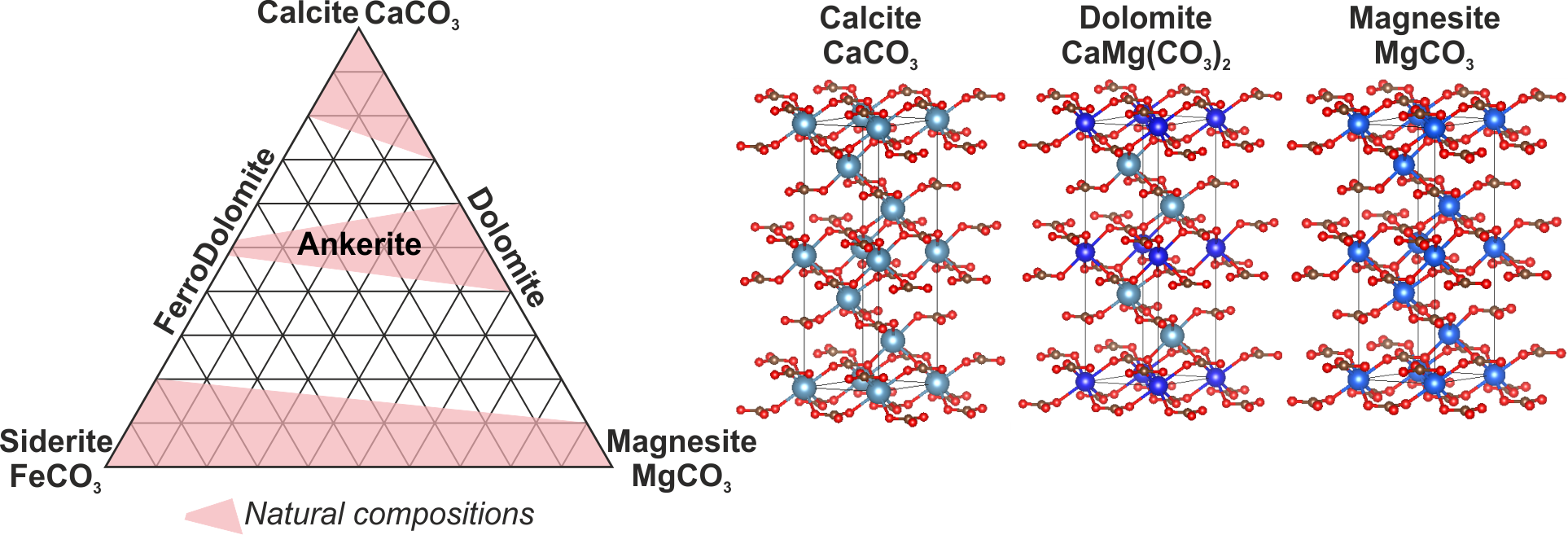}
    \caption{Ternary plot of the calcite – magnesite – siderite mineral series with highlighted naturally occurring compositions of carbonate mineral species taken from \cite{Deer2013} together with visualization of the calculated crystal structure of calcite, dolomite and magnesite used for magnetic properties calculations.}
    \label{fig:intro}
\end{figure}

The calcite-magnesite series is part of the larger calcite-magnesite system, characterized by limited cation exchange in naturally occurring mineral species (Fig.~\ref{fig:intro}). The magnetic properties of carbonate minerals have been of interest in Earth sciences since the 19th century (\cite{Tyndall1851}), with individual mineral species investigated using both low-field and high-field experimental methods (\cite{Konig1887}, \cite{Voigt1907}, \cite{Owens1978}, \cite{Borradaile1981}, \cite{Rochette1988}, \cite{Schmidt2006}, \cite{Schmidt2007}). Among these, calcite is the most extensively studied mineral, with the anisotropy of magnetic properties being the best characterised \cite{Schmidt2006}. In contrast, magnesite and dolomite have been studied less extensively, and the results remain ambiguous and inconclusive \cite{Voigt1907}, \cite{Schmidt2007}. A significant limitation in these studies is the insufficient or absent reporting of the chemical composition of the analysed samples. For this study, data from \cite{Schmidt2006, Schmidt2007}, complemented by analyses using LA-ICP-MS (Laser Ablation Inductively Coupled Plasma Mass Spectrometry), have been identified as the most reliable. The experimentally determined magnetic properties of calcite, magnesite and dolomite are summarised in Tab.~\ref{tab1}.

Calcite (CaCO$_3$) exhibits trigonal symmetry of the R$\overline{3}$c space group (No.~167). Its structure is characterized by a face-centred rhombohedral unit cell. Various divalent cations may partially substitute for calcium (Ca$^{2+}$)  in calcite, with the most common being magnesium (Mg$^{2+}$), forming a solid solution of MgCO$_3$ in calcite. Other frequent substitutions involve manganese (Mn$^{2+}$) and iron (Fe$^{2+}$) \cite{Deer2013}.
Following the usual AMS convention, $k_1\geq k_2\geq k_3$ denote the principal values (eigenvalues) of the susceptibility tensor, taken with their sign, and the corresponding eigenvectors are the principal susceptibility axes. In the crystal coordinate system used in this paper ($z\parallel c$), the susceptibility tensor of the trigonal carbonates is diagonal, with $\chi_{xx}=\chi_{yy}$ perpendicular to the $c$-axis and $\chi_{zz}$ parallel to it. Hence, $k_i$ and $\chi_{\alpha\alpha}$ denote the same quantities, ordered by magnitude and labelled by direction, respectively.
Pure calcite is diamagnetic, with the minimum (most negative) susceptibility $k_3$, whose principal axis is oriented parallel to the crystallographic $c$-axis, while the principal axes of $k_1=k_2$ lie within the plane of the crystal $a$-axes (some studies use absolute values of the diamagnetic susceptibility, in which case the $c$-axis appears as the maximum susceptibility axis). The AMS tensor for a single crystal of pure calcite, as defined by \cite{Schmidt2006}, has the following values $k_1 = k_2 = -11.723 \times 10^{-6}$\,[SI], $k_3 = -12.823 \times 10^{-6}$\,[SI], and a mean susceptibility of $k_m = -12.08 \times 10^{-6}$\,[SI]. 
Expressed as the induced magnetic moment per applied magnetic field and per formula unit, these susceptibilities are $k_3=-6.748\times10^{-5}$\,$\mu_B$\,T$^{-1}$\,f.u.$^{-1}$\ and $k_1=k_2=-6.169\times10^{-5}$\,$\mu_B$\,T$^{-1}$\,f.u.$^{-1}$\, where f.u.\ (formula unit) denotes CaCO$_3$.
The empirical relationships describing  the paramagnetic susceptibility difference and the increase in susceptibility with rising iron (Fe) concentration in calcite are expressed as:
$ \Delta k_\mathrm{para}$\,[m$^3$/kg] = Fe content [ppm] $\times (1\pm0.1)\times10^{-12}$\,[m$^3$/kg/ppm], where $\Delta k=k_1-k_3$.
The susceptibility increases by approximately  $2.3\times 10^{-10}$ [m$^3$/kg] for every 100\,ppm Fe \cite{Schmidt2006}. Iron-rich calcite with paramagnetic properties exhibits an “inverse” magnetic fabric characterized by a prolate ellipsoid, where the $k_1$ axis aligns with the c-axis of the crystal lattice. The transition between oblate and prolate ellipsoids occurs at about 400\,ppm of Fe$^{2+}$ within the calcite crystal lattice. Additionally, \cite{Rochette1988} describes an empirical relationship that outlines how magnetic susceptibility changes with increasing temperature, and \cite{Schmidt2007} showed that the paramagnetic $\Delta k$ of Fe-bearing calcite increases 13.3 times upon cooling from room temperature to 77\,K, independently of the Fe content.

Magnesite (MgCO$_3$) has a structure and symmetry similar to calcite, but with a smaller unit cell due to the smaller size of Mg$^{2+}$ ions. A complete solid solution series exists between magnesite and siderite (FeCO$_3$), while limited solubility of CaCO$_3$ in magnesite has been observed \cite{Deer2013}.
Reported susceptibility values for magnesite are limited and inconsistent. Historically, values for $k_m$ ranged from $28 \times 10^{-6}$ to $137 \times 10^{-6}$\,[SI]  \cite{Bleil1982}
Diamagnetic susceptibility value $-5.01 \times 10^{–6}$\,[SI] was reported by \cite{Schmidt2007} for sample Mg1, which contains 440\,ppm Fe and 750\,ppm Mn. A weak,  oblate AMS tensor with $k_3$ aligned parallel to the $c$-axis is proposed, with a shift to a prolate shape as $k_1$ aligns with the $c$-axis upon increasing  Fe$^{2+}$ content.

Dolomite Ca$_{0.5}$Mg$_{0.5}$CO$_3$ has a structure with slightly lower symmetry than calcite, characterised by the R$\overline{3}$ space group (No.~148). Its structure can be visualised as a combination of layers of calcite and magnesite. Continuous replacement of Mg$^{2+}$ by Fe$^{2+}$ occurs in dolomite, forming a solid solution series from ankerite Ca$_{0.5}$(Fe, Mg)$_{0.5}$CO$_3$ to ferro-dolomite Ca$_{0.5}$Fe$_{0.5}$CO$_3$. In the crystal structure,  Mg$^{2+}$ is commonly replaced by Mn$^{2+}$ \cite{Deer2013}.
Pure dolomite is expected to be diamagnetic, exhibiting an oblate AMS ellipsoid shape with $k_3$ aligned parallel to the $c$-axis and becoming prolate with $k_1$ parallel to the $c$-axis with increasing Fe$^{2+}$ content. Reported susceptibility and $\Delta k$ values for the purest samples in the dataset range from $–9.78\times10^{–6}$ to $-6.46\times 10^{–6}$\,[SI] and from $0.76\times 10^{-6}$ to $1.28\times 10^{-6}$\,[SI], respectively. These values are strongly influenced by Fe$^{2+}$ and Mn$^{2+}$ content \cite{Schmidt2007}.

\begin{table}[]
    \centering
\begin{tabular}{lllll}
& $\chi_{\mathrm{trace}}$ [$10^{-6}$\,SI]
& $\chi_{xx}$ [$10^{-6}$\,SI]
& $\chi_{zz}$ [$10^{-6}$\,SI]
& $\chi_{xx}/\chi_{zz}$
\\ \hline
\textbf{Calcite:} & & & &
\\
\cite{Konig1887}
& -14.78 
& -14.30 
& -15.73 
& 0.909 
\\
\cite{Voigt1907}
& -12.87   
& -12.4  
& -13.8  
& 0.899 
\\
\cite{Owens1978}
& -12.9 
& -12.50 
& -13.69 
& 0.913 
\\
\cite{Rochette1988}
& -12.57 
& -12.0  
& -13.7  
& 0.876  
\\
\cite{Schmidt2006}
& -12.08 
& -11.72 
& -12.82
& 0.914 
\\ \hline
\textbf{Magnesite:}    & & &  & \\
\cite{Schmidt2007}, sample Mg1 
& -5.01
& -4.77
& -5.19 
& 0.918 
\\ \hline
\textbf{Dolomite:}    & & & & 
\\
\cite{Voigt1907}
& 36.16 
& 28.37 
& 43.94 
& 0.646 
\\
\cite{Schmidt2007}, sample D2    
& -6.46 
& -5.85 
& -7.13 
& 0.82  
\\
\cite{Schmidt2007}, sample D3    
& -9.78 
& -9.29
& -10.05 
& 0.924 
\\ \hline
\end{tabular}
    \caption{Experimentally determined anisotropy of magnetic properties of calcite, magnesite and dolomite. For dolomite, two samples are selected from \cite{Schmidt2007} with the lowest Fe content. Fe and Mn contents (mass concentrations) of the samples of \cite{Schmidt2007}: magnesite Mg1 440\,ppm Fe and 750\,ppm Mn; dolomite D2 56\,ppm Fe and 460\,ppm Mn; dolomite D3 74\,ppm Fe and 191\,ppm Mn.}
    \label{tab1}
\end{table}

\section{Methods}

\subsection{Ab-initio calculation of the diamagnetic susceptibility}
\label{sec:methods}

The diamagnetic susceptibility of calcite was computed from first principles using density functional theory (DFT) with the augmented plane-wave (APW) method as implemented in the Wien2k code \cite{Blaha2020}.
Ground-state electronic structure is calculated using
experimental lattice parameters and experimental internal atomic positions
of calcite, dolomite and magnesite. 
Muffin-tin radii are chosen to maximise their respective radii. 
Exchange-correlation potentials are LDA, PBE
and mBJ, where the mBJ potential is a phenomenological modification of the PBE approach adjusted to predict experimental gap size \cite{Tran2009}.
The number of basis wavefunctions is defined by $\mathrm{RMT}_\mathrm{min}G_\mathrm{max}=6.0$, where $\mathrm{RMT}_\mathrm{min}$ is the minimal muffin-tin radius and $G_\mathrm{max}$ is the maximal length of $\vec{G}$-vector used in basis construction. Brillouin zone sampling was $8\times8\times8$. We have verified that this setting yields converged susceptibility values (the difference in susceptibility was below 0.2\% compared with the $\vec{k}$-mesh $17\times17\times17$ and $\mathrm{RMT}_\mathrm{min}G_\mathrm{max}=7.0$). Convergence criteria for total energy and charge were $5\times 10^{-6}$\,Ry and $5\times 10^{-6}$, respectively. Core states were treated fully relativistically. The calculated magnetic susceptibility tensor agrees for calculations with either spin-orbit coupling included or excluded. Hence, the spin-orbit coupling is omitted in the calculation.
Used experimental room-temperature lattice constants are: calcite ($a=4.9900$\,\AA, $c=17.0615$\,\AA), dolomite
($a=4.8033$\,\AA, $c=15.9840$\,\AA) and magnesite ($a=4.6370$\,\AA, $c=15.0230$\,\AA).

Fe- and Mn-doped carbonates were modelled by supercells built from the 30-atom hexagonal unit cell (six formula units). One cation of the $1\times1\times1$, $1\times1\times2$, $2\times2\times1$ and $2\times2\times2$ hexagonal supercells (30, 60, 120 and 240 atoms) is replaced by the dopant, giving atomic concentrations $x=\frac{1}{6}$, $\frac{1}{12}$, $\frac{1}{24}$ and $\frac{1}{48}$ per cation site. The dopant substitutes Ca in calcite and Mg in dolomite and magnesite, i.e.\ it is octahedrally coordinated by six oxygen atoms. The lattice constants were scaled linearly with $x$ between those of the host and siderite (FeCO$_3$), and the internal atomic positions were not relaxed. The shortest distance between a dopant and its periodic images is the lattice constant $a$ for $x=\frac{1}{6}$ and $\frac{1}{12}$ (4.94--4.97\,\AA\ in calcite, about 4.8\,\AA\ in dolomite and 4.65\,\AA\ in magnesite) and $2a$ for $x=\frac{1}{24}$ and $\frac{1}{48}$ (9.95--9.97\,\AA\ in calcite, 9.6\,\AA\ in dolomite and 9.3\,\AA\ in magnesite). The calculations are spin polarised, with the dopant spins of all periodic images aligned parallel. The Brillouin zone was sampled by 784, 512, 256 and 128 $\vec{k}$-points for $x=\frac{1}{6}$, $\frac{1}{12}$, $\frac{1}{24}$ and $\frac{1}{48}$, respectively.

\subsection{Formalism for Diamagnetic Susceptibility}
The diamagnetic susceptibility tensor arises from the second-order energy variation due to a static uniform magnetic field $\vec{B}$. 
\begin{equation}
\chi_{\alpha\beta}=
\left.
-\frac{1}{V} \frac{\partial^2 E}{\partial B_\alpha \partial B_\beta} 
\right|_{B=0}  
\end{equation}
where $\alpha$, $\beta$ are $x$, $y$, $z$ coordinates. However, as changes of the total energy $E$ due to the external magnetic field $\vec{B}$ are tiny, the susceptibility is calculated using perturbation theory. First, the current density $\vec{j}(\vec{r})$ of each electron inside the unit cell is calculated using linear response theory to the external magnetic field $\vec{B}$. Theory shows that those current densities consist of the paramagnetic current density $\vec{j}_\mathrm{para}$ (given by the first-order perturbation of the wavefunction and first-order perturbation in $\vec{A}$ of the current operator) and diamagnetic current density $\vec{j}_\mathrm{dia}$ (given by the second-order perturbation in $\vec{A}$ of the current operator),
\begin{align}
\vec{j}_\mathrm{ind}({\vec{r}}) 
&= -\frac{\delta H}{\delta{\vec{A}}({\vec{r}})} = \vec{j}_\mathrm{para}({\vec{r}}) + \vec{j}_\mathrm{dia}({\vec{r}})
\\
{\vec{j}}_{para}({\vec{r}}) 
&= - 2\mathcal{R}\mathrm{e}\left[
f(E_{n{\vec{k}}})\ \frac{\braket{n{\vec{k}}|\vec{J}_\mathrm{para}({\vec{r}})|m{\vec{k}}} \braket{m{\vec{k}}|H_{1}|n{\vec{k}}}}{E_{n{\vec{k}}} - E_{m{\vec{k}}}}
\right]
\\
{\vec{j}}_\mathrm{dia}({\vec{r}}) 
&= f(E_{n{\vec{k}}})\ \braket{n{\vec{k}}|J_\mathrm{dia}(\vec{r})|n{\vec{k}}}
\end{align}
where
$H_{1}=\frac{e}{2m}(\vec{r} \times \vec{p})\cdot \vec{B}$
is the perturbation Hamiltonian in the first order in the vector
potential $\vec{A}(\vec{r})$, which was selected in the form
$\vec{A}(\vec{r}) = \frac{1}{2}\vec{B} \times \vec{r}$.
The paramagnetic and diamagnetic current operators are
\begin{align}
\vec{J}_\mathrm{para}(\vec{r}) 
&= 
-\frac{e}{2m}\lbrack \vec{p}\ket{\vec{r}}\bra{\vec{r}} + \ket{\vec{r}} \bra{\vec{r}} \vec{p} \rbrack
\\
\vec{J}_\mathrm{dia}(\vec{r}) &= \frac{e^{2}}{m}{\vec{A}}(\vec{r})  
\ket{\vec{r}}\bra{\vec{r}}
= 
\frac{e^{2}}{2m}\vec{B} \times \vec{r} 
\ket{\vec{r}}\bra{\vec{r}}
\end{align}
where $\ket{n\vec{k}}$ is the Bloch state
for band $n$ at wavevector $\vec{k}$ and
$f(E_{n\vec{k}})$ is the occupation probability. The
summation over quantum numbers $n\vec{k}$ includes all occupied states in the
Brillouin zone. Knowing the spatial distribution of current density
$\vec{j}_\mathrm{ind}(\vec{r})$
inside the unit cell, the magnetic
field in an arbitrary position $\vec{R}$ in the crystal is
provided by the Biot-Savart law
\begin{equation}
\vec{B}_\mathrm{ind}(\vec{R}) = \frac{\mu_{0}}{4\pi}
\int \mathrm{d}^{3}\vec{r}\,\,
\vec{j}_\mathrm{ind}(\vec{r}) \times 
\frac{\vec{r}-\vec{R}}{|\vec{r}-\vec{R}|^{3}}.
\end{equation}
The diamagnetic magnetization 
$\vec{M} = \frac{1}{\mu_{0}}\langle\vec{B}_\mathrm{ind}\rangle$ is the average of the induced internal magnetic fields 
$\vec{B}_\mathrm{ind}(\vec{R})$,
providing the magnetic susceptibility tensor
\begin{equation}
\chi_{\alpha\beta} = \frac{\mu_{0}M_\beta}{B_\alpha}.
\end{equation}

When the magnetic field is expressed in individual atomic positions, it
provides nuclear magnetic resonance (NMR) screening. The implementation
follows the gauge-invariant formulation described in 
\cite{Laskowski2012, Laskowski2014}.
The calculations were performed in the Wien2k code
\cite{Blaha2020}, and its NMR package.

\subsection{Diamagnetic susceptibility}
\label{sec:dia}

The calculated diagonal values of the magnetic susceptibility are shown
in Table~\ref{tab2} for calcite, magnesite, and dolomite. In the case of calcite,
the agreement between experimental and calculated susceptibility is
excellent. Using LDA and PBE exchange correlation potential, the
calculated susceptibility is underestimated. On the other hand, using
mBJ exchange correlation, the susceptibility is slightly overestimated.
It suggests that the correct susceptibility for dolomite and magnesite
is between the values calculated for PBE and mBJ. 
The raw experimental values for dolomite and magnesite are considerably less negative than the calculated ones. This is not a failure of the calculations: the measured samples contain Fe and Mn (Tab.~\ref{tab1}), whose paramagnetic (Curie) contribution is positive and, at room temperature, comparable to the diamagnetic susceptibility itself. We subtracted this contribution using the measured Fe and Mn contents of each sample and the Curie model of Sec.~\ref{sec:spinpara} (paramagnetic anisotropy of Fe as in Sec.~\ref{sec:orbital}, scaled by the volume per formula unit). The corrected experimental values (rows `exp., corr.' in Tab.~\ref{tab2}) are $-15.5\times10^{-6}$\,[SI] for magnesite and $-12.1\times10^{-6}$\,[SI] for dolomite D3 ($-11.2\times10^{-6}$\,[SI] for D2), i.e.\ within about $1\times10^{-6}$\,[SI] of the calculated values. The corrected anisotropy of magnesite, $\chi_{zz}-\chi_{xx}=-1.74\times10^{-6}$\,[SI], agrees with the PBE value $-1.73\times10^{-6}$\,[SI]. The remaining differences are comparable to the spread between the corrected values of individual dolomite samples (from $-11.2$ to $-13.2\times10^{-6}$\,[SI], \ref{app:residuals}).

\begin{figure}
\centering
\includegraphics[width=0.45\textwidth]{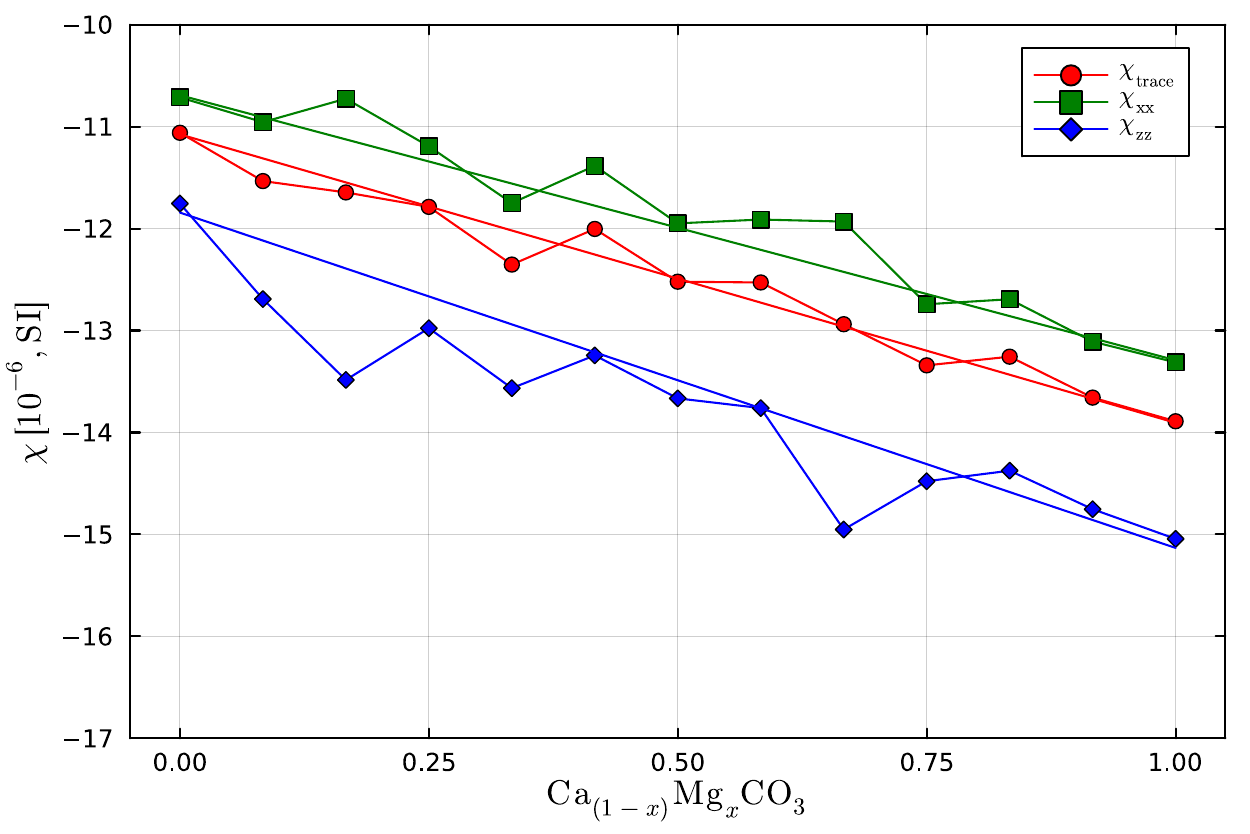}
\caption{Calculated susceptibility for Ca\textsubscript{1-x}Mg\textsubscript{x}CO\textsubscript{3}, where $x=0$, $0.5$, $1$ corresponds to calcite, dolomite and magnesite. The dependence was calculated using the supercell approach and PBE exchange-correlation potential.}
\label{fig:dia-cal-mag}
\end{figure}

Fig.~\ref{fig:dia-cal-mag} 
shows the calculated dependence of magnetic susceptibility for
Ca\textsubscript{1-x}Mg\textsubscript{x}CO\textsubscript{3}, where $x=0$,
$0.5$, $1$ being calcite, dolomite and magnesite, respectively. The dependence was calculated using the supercell approach and PBE exchange-correlation potential. The values of susceptibility are scattered due to the supercell approach, which keeps artificial
periodicity, leading to the `noise'. However, one can see a linear
decrease of susceptibility (i.e.\ increase of its absolute value) with
increasing Mg concentration. This is due to a reduction of the unit cell
volume (reduction of lattice constants). Namely, when going from
CaCO\textsubscript{3} to MgCO\textsubscript{3}, the volume of the unit
cell decreases by 24\%, and trace (mean) susceptibility increases by
27\%, demonstrating that the increase in susceptibility is dominantly
due to the reduction of the volume of the unit cell. Surprisingly, the
reduced number of electrons from Ca (20$\overline{e}$) to Mg (12$\overline{e}$) does not vary the
susceptibility value, as it is the number of core electrons which
varies, and the susceptibility of core electrons (per electron) is more
than one order smaller compared to the susceptibility contribution of
valence electrons. Therefore, it is the number of valence electrons and
the unit cell volume that determine the strength of the magnetic
susceptibility.

\begin{table}[]
\begin{tabular}{llllll}
    & $\chi_\mathrm{trace}$ [$10^{-6}$\,SI] 
    & $\chi_{xx}$ [$10^{-6}$\,SI] 
    & $\chi_{zz}$ [$10^{-6}$\,SI] 
    & $\chi_{xx}/\chi_{zz}$ 
    & $\chi_{zz}-\chi_{xx}$ [$10^{-6}$\,SI]
    \\ \hline
\multicolumn{6}{l}{\textbf{Calcite:}} 
\\
exp & -12.09  & -11.71 & -12.86 & 0.911 & -1.15
\\ 
LDA & -10.97 & -10.65  & -11.61  & 0.917 & -0.96
\\ 
PBE & -11.06 & -10.71 & -11.75  & 0.911 & -1.04
\\ 
mBJ & -12.22  & -11.93  & -12.81  & 0.931 & -0.88
\\ \hline
\multicolumn{6}{l}{\textbf{Magnesite:}} 
\\ 
exp & -5.01  & -4.77  & -5.19 & 0.918   & -0.42
\\
exp., corr. & -15.50 & -14.82 & -16.56 & 0.895 & -1.74
\\
LDA & -13.93 & -13.36  & -15.07  & 0.886 & -1.71
\\
PBE & -13.89  & -13.31 & -15.04 & 0.884  & -1.73
\\
mBJ & -14.76  & -14.39 & -15.52  & 0.927 & -1.23
\\ \hline
\multicolumn{6}{l}{\textbf{Dolomite:}} 
\\
exp & -9.78  & -9.29 & -10.05 & 0.924  & -0.76
\\
exp., corr. & -12.09 & -11.53 & -12.50 & 0.922 & -0.97
\\
LDA & -12.46 & -11.90  & -13.57  & 0.877  & -1.67
\\
PBE & -12.52 & -11.95  & -13.67  & 0.874  & -1.72
\\
mBJ & -13.59  & -13.17 & -14.44 & 0.911  & -1.27  
\\ \hline
\end{tabular}
\caption{Ab-initio calculated and experimentally determined (volume) magnetic susceptibility of calcite, dolomite and magnesite. Experimental values: calcite \cite{Schmidt2006}; magnesite sample Mg1 and dolomite sample D3 \cite{Schmidt2007} (Tab.~\ref{tab1}). Rows `exp., corr.' give the experimental values after subtraction of the paramagnetic contribution of the measured Fe and Mn contents (Sec.~\ref{sec:dia}).}
\label{tab2}
\end{table}

\subsection{Anisotropy of diamagnetic susceptibility}

Calcite, dolomite, and magnesite have a strong anisotropy of magnetic susceptibility (AMS). This anisotropy can be expressed as the ratio $\chi_{xx}/\chi_{zz}\approx 0.9$, i.e.\ susceptibility (induced magnetic field) is larger along the $c$-axis, in agreement between experiment and ab-initio calculations. Another way to express AMS is difference $\Delta \chi^\mathrm{(SI)}=\chi_{zz}-\chi_{xx}$. At zero Fe concentration, the experimental value is \cite{Schmidt2006}, Tab.~\ref{tab1}
\begin{equation}
\Delta\chi^{\mathrm{(SI)}} = \chi_{zz}^{\mathrm{(SI)}} - \chi_{xx}^{\mathrm{(SI)}} = (-1.10
\pm 0.01)\,\,  \mathrm{[10^{-6}, SI]}
\label{eq:dia-aniso-exp}
\end{equation}
which agrees well with the PBE ab-initio calculated AMS, being $-1.04$\, [$10^{-6}$, SI] (Tab.~\ref{tab2}).

The origin of the AMS is due to different orbital motion of electrons in the $x-y$ plane (which generates a larger magnetic moment) compared to the circular motion in the $x-z$ or $y-z$ planes (which generates a smaller magnetic moment) at a given external magnetic field. 

 We stress out that the AMS does not originate from spin-orbit coupling, as the same values of AMS are ab-initio calculated when spin-orbit coupling is included or excluded.

\subsection{Doping calcite by Fe and Mn: spin paramagnetisation}
\label{sec:spinpara}

\cite{Schmidt2006} measured at room temperature how the mass susceptibility of calcite increases with Fe and Mn content: by $2.3\times10^{-10}$\,m$^3$/kg per 100\,ppm Fe and by $3.4\times10^{-10}$\,m$^3$/kg per 100\,ppm Mn, where ppm denotes mass concentration. We describe the paramagnetic contribution of dilute, non-interacting ions by the Curie law (the Curie--Weiss model with zero Weiss temperature) for the volume susceptibility
\begin{equation}
\chi^\mathrm{(SI)}=\frac{\mu_0 N_A n_\mathrm{Fe} (\mu_\mathrm{eff}\mu_B)^2}{3k_B T}
\label{eq:Curie}
\end{equation}
where $n_\mathrm{Fe}$ is the molar density of Fe ions and the effective magnetic moment, in units of $\mu_B$, is (for details see \ref{app:curie})
\begin{equation}
\mu_\mathrm{eff}= g\sqrt{S(S+1)}.
\label{eq:mueff}
\end{equation}
The experimentally observed slope with Fe and Mn concentration corresponds to the experimental value of the effective magnetic moment of a single atom in the calcite matrix being $\mu_\mathrm{eff, Fe}=4.94$\,$\mu_B$ for Fe ions and $\mu_\mathrm{eff, Mn}=5.95$\,$\mu_B$ for Mn ions (details of unit conversion see Eq.~(\ref{eq:conv:mueff_w})).

We ab-initio calculated the magnetic moment of Fe and Mn inclusions into the calcite matrix, assuming Fe and Mn atoms replacing Ca atoms. The calculations were based on the supercell approach (Sec.~\ref{sec:methods}), with atomic concentrations of Fe or Mn of $\frac{1}{6}$, $\frac{1}{12}$, and $\frac{1}{24}$ (and $\frac{1}{48}$ for Fe), using the PBE exchange-correlation potential. In the case of Fe, we also calculated mBJ and PBE+U (with U$_\mathrm{Fe}$ = 4.4--5.5\,eV). In all those exchange-correlation models, the magnetic moment of a single Fe atom was $\mu_{\mathrm{Fe}}=4.00$\,$\mu_B$, corresponding to spin state 3d$^6$, i.e.\ five spins-up and one spin-down, providing total spin $S=2$ (corresponding to a high-spin Fe$^{2+}$ ion). Substituting $S=2$ into the relation for the effective magnetic moment, Eq.~(\ref{eq:mueff}), we got $\mu_\mathrm{eff, Fe}=4.90\,\mu_B$, in perfect agreement with the experiment.
Similarly, the magnetic moment of Mn is calculated to be in the state of a high-spin Mn$^{2+}$ ion (3d$^5$ configuration, $S=5/2$), i.e.\ five electrons having spin-up and zero electrons having spin-down. Assuming $g=2$, the corresponding effective magnetisation is $\mu_\mathrm{eff, Mn}=5.92\,\mu_B$, in excellent agreement with the experiment.

\begin{figure}
\centering
\includegraphics[width=0.95\linewidth]{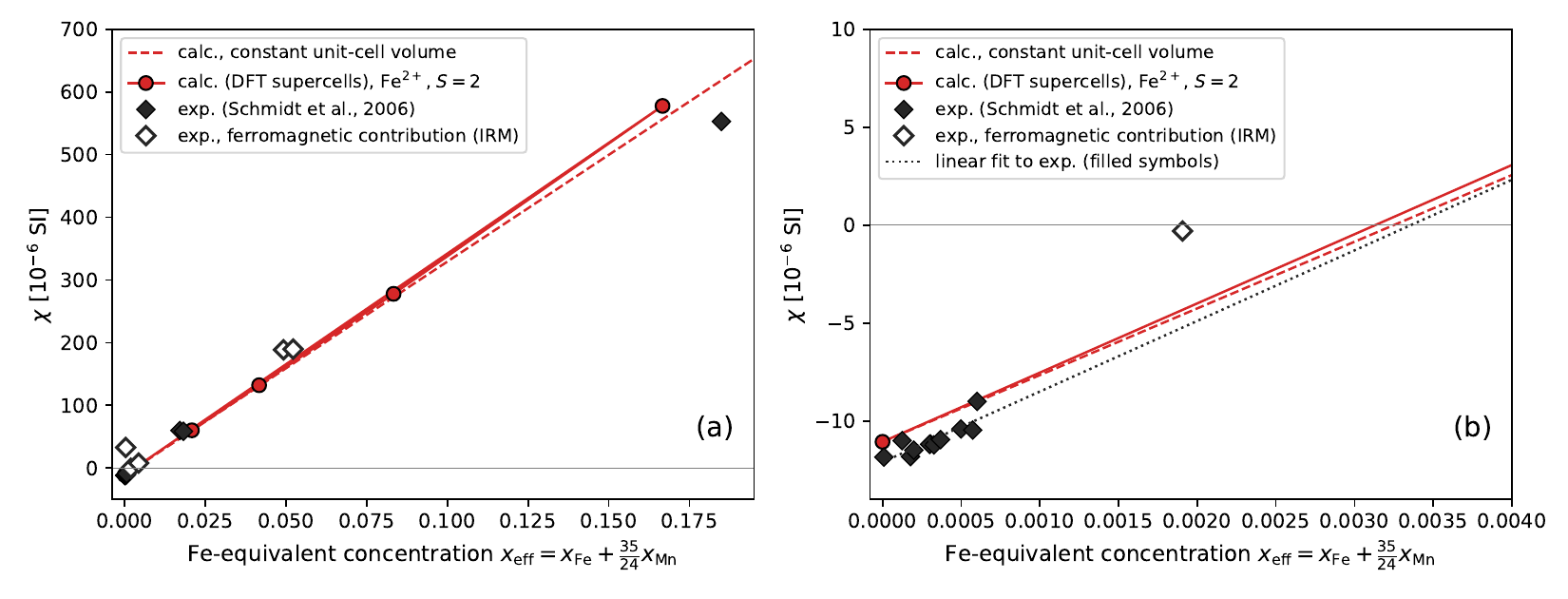}
\caption{
(a) Magnetic susceptibility $\chi$ of Fe-doped calcite Ca$_{1-x}$Fe$_{x}$CO$_3$ versus the Fe-equivalent concentration $x_\mathrm{eff}$, Eq.~(\ref{eq:xeff}). Red circles: total (diamagnetic and paramagnetic) susceptibility calculated for $x=0$, $\frac{1}{48}$, $\frac{1}{24}$, $\frac{1}{12}$ and $\frac{1}{6}$, with lattice constants scaled linearly with $x$ between calcite and siderite. The paramagnetic (Curie) part is evaluated at ${T=300}$\,K, the diamagnetic part does not depend on temperature. Red dashed line: the same, assuming a constant unit-cell volume of calcite. Black diamonds: experimental data measured at room temperature \cite{Schmidt2006}, with $x_\mathrm{eff}$ obtained from the measured Fe and Mn contents of each sample; open diamonds mark samples with a ferromagnetic contribution detected by IRM. (b) Zoom of (a) into the low-concentration range. Dotted line: linear fit to the filled diamonds.
}
\label{fig:xicalTdep}
\end{figure}

Using this model, Fig.~\ref{fig:xicalTdep} shows the calculated total (diamagnetic and paramagnetic) susceptibility of Fe-doped calcite, with the paramagnetic part evaluated at 300\,K, and compares it with the room-temperature experimental data of \cite{Schmidt2006}. The experimental samples contain both Fe and Mn. Since the Curie contributions of different ions add, each weighted by $\mu_\mathrm{eff}^2$ (\ref{app:curie}), the samples are compared with the Fe-only calculation using the Fe-equivalent concentration
\begin{equation}
x_\mathrm{eff} = x_\mathrm{Fe} + \frac{\mu^2_\mathrm{eff,Mn}}{\mu^2_\mathrm{eff,Fe}}\, x_\mathrm{Mn} = x_\mathrm{Fe} + \frac{35}{24}\, x_\mathrm{Mn},
\label{eq:xeff}
\end{equation}
where $x_\mathrm{Fe}$ and $x_\mathrm{Mn}$ are atomic concentrations per cation site obtained from the measured mass concentrations (\ref{app:conc}), and $\mu^2_\mathrm{eff}=g^2S(S+1)\,\mu_B^2$ is $24\,\mu_B^2$ for Fe$^{2+}$ and $35\,\mu_B^2$ for Mn$^{2+}$.

The calculated susceptibility depends linearly on $x_\mathrm{eff}$, with a slope of $3404\times10^{-6}$\,[SI] per unit $x_\mathrm{eff}$ at 300\,K. The small difference between the supercell points and the dashed line is due to the decrease of the unit-cell volume with Fe content; it is not of numerical origin. At low concentrations ($x_\mathrm{eff}<0.003$, Fig.~\ref{fig:xicalTdep}b), a linear fit to the experimental samples without ferromagnetic contribution gives $\chi = -12.10\times10^{-6} + 3603\times10^{-6}\,x_\mathrm{eff}$\,[SI]. Its intercept equals the intrinsic diamagnetic susceptibility of calcite \cite{Schmidt2006}, and its slope is within 6\% of the calculated one. The samples lying above the calculated line (open symbols in Fig.~\ref{fig:xicalTdep}) are those in which \cite{Schmidt2006} detected a ferromagnetic contribution by isothermal remanent magnetisation (IRM), i.e.\ part of their Fe is not substituted in the calcite lattice. Larger deviations appear only at high concentrations: the Mn-rich sample C14 ($x_\mathrm{Mn}=0.127$) lies 10\% below the calculated line, which is consistent with antiferromagnetic interactions between neighbouring Mn ions. The differences between measured and calculated susceptibility for all samples are shown in \ref{app:residuals}.

\begin{figure}
\centering
\includegraphics[width=0.95\linewidth]{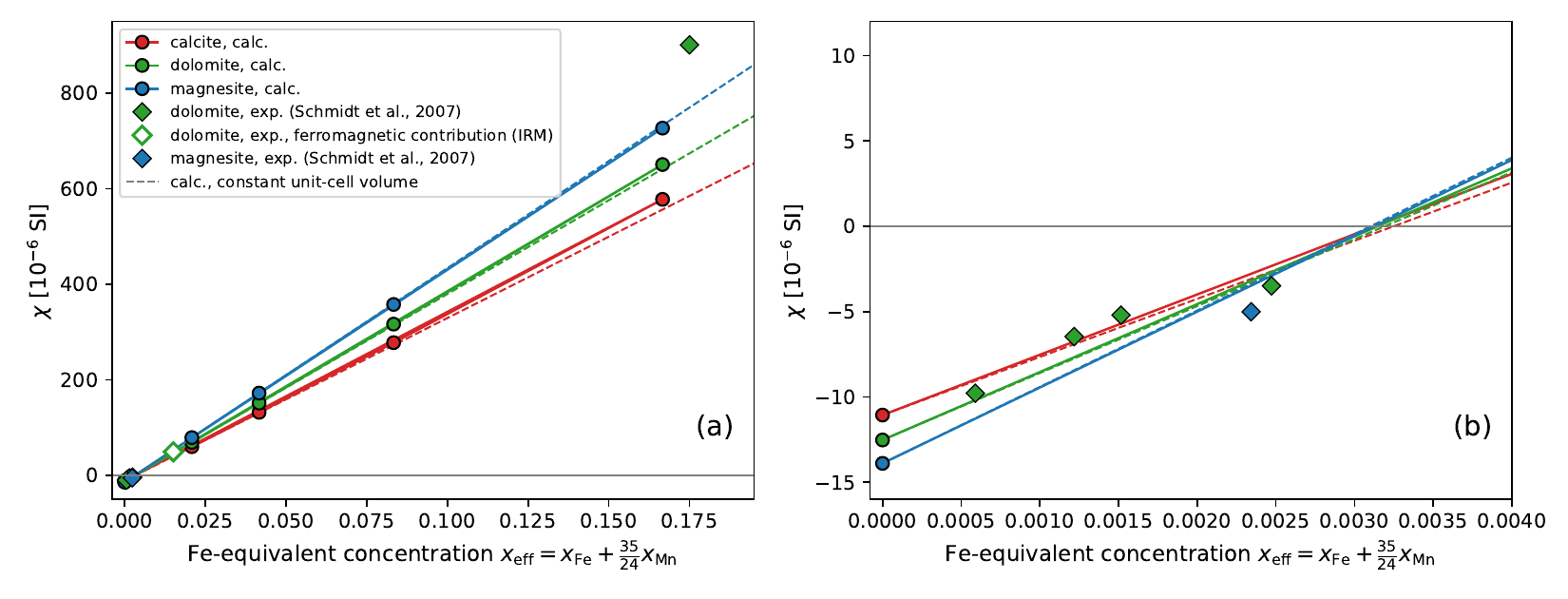}
\caption{(a) Calculated total susceptibility (paramagnetic part at ${T=300}$\,K) of Fe-doped calcite (red), dolomite (green) and magnesite (blue) versus Fe-equivalent concentration $x_\mathrm{eff}$, Eq.~(\ref{eq:xeff}); for the calculations, $x_\mathrm{eff}=x_\mathrm{Fe}$. Circles with full lines: supercell calculations with lattice constants scaled linearly with $x_\mathrm{Fe}$ between the host and siderite. Dashed lines: constant unit-cell volume of the host. Diamonds: experimental data for dolomite and magnesite measured at room temperature \cite{Schmidt2007}; the open diamond marks the dolomite sample with a ferromagnetic contribution detected by IRM. (b) Zoom of (a) into the low-concentration range.
}
\label{fig:xiTdep}
\end{figure}

Fig.~\ref{fig:xiTdep} shows the calculated dependence for calcite, dolomite and magnesite, together with the room-temperature experimental data for dolomite and magnesite \cite{Schmidt2007}. The low-concentration dolomite samples and the magnesite sample lie close to the calculated lines. The Fe-rich dolomite D4 (9.6\,wt\% Fe) lies 34\% above the calculated line, which may indicate an orbital contribution to the effective moment of Fe$^{2+}$ beyond the spin-only value or Fe-bearing inclusions. Finally, we note that the calculated curves contain only the spin (Curie) contribution of the dopants; the orbital contribution of Fe discussed in Sec.~\ref{sec:orbital} adds a small, positive, temperature-independent term (about 20\% of the Curie slope at 300\,K in the $x=\frac{1}{6}$ supercell).

\subsection{Anisotropy paramagnetic susceptibility: orbital magnetisation contribution}
\label{sec:orbital}

\cite{Schmidt2006} experimentally determined the empirical relationship of paramagnetic mass susceptibility anisotropy $\Delta\chi=\chi_{zz}-\chi_{xx}$ to be 
\begin{equation}
\Delta\chi_\mathrm{mass, para} \mathrm{[m^3/kg]} = w_\mathrm{Fe}\ [\mathrm{mass\ concentration \ ppm}] \times (+1\pm0.1) \times 10^{-12}\ [\mathrm{m}^3\mathrm{/kg/ppm}], 
\end{equation}
where $\Delta\chi=\chi_{zz}-\chi_{xx}=k_1-k_3=\Delta k$, as the maximum susceptibility $k_1$ of Fe-bearing calcite is parallel to the $c$-axis,
which can be converted to anisotropy volume susceptibility 
per atomic concentration $x_\mathrm{Fe}$ (details of the conversion Eq.~(\ref{eq:conv:Deltachi_xFe}))
\begin{equation}
\frac{\Delta \chi^{\mathrm{(SI)}}_\mathrm{vol, para}}{x_\mathrm{Fe}} 
= 1512\,\, [10^{-6}, \mathrm{SI}]
\label{eq:Deltachi_xFe}
\end{equation}

In order to express the contribution of Fe doping beyond the Curie-Weiss model and spin-magnetisation, we have ab initio expressed orbital magnetisation (induced internal magnetic field) induced by the external magnetic field. 
Namely, using supercell approches, we calculated Fe doping $x_\mathrm{Fe}=\frac{1}{6}$, i.e.\ for structure FeCa$_5$C$_6$O$_{18}$. For this supercell, we calculated magnetic susceptibility and its AMS. These results are preliminary. Owing to the very high computational cost of the spin-polarised orbital-susceptibility calculation, it was feasible only for the smallest supercell, and its convergence with respect to the $\vec{k}$-point sampling could not be fully verified. The numbers below should therefore be regarded as estimates, and new, more efficient methods are needed for a quantitative description (Sec.~\ref{sec:discussion}).

As the ground state of Fe-doped calcite is ferromagnetic, the (volume) susceptibility was calculated separately for spin-up and spin-down electrons 
\begin{align}
\chi_{xx} &= \chi_{xx}^\uparrow +  \chi_{xx}^\downarrow = 27.38 +7.26 = 34.64 \,\,\mathrm{[10^{-6}, SI]}
\\
\chi_{zz} &= \chi_{zz}^\uparrow +  \chi_{zz}^\downarrow = 282.78 +5.54 = 288.32\,\, \mathrm{[10^{-6}, SI]}
\end{align}
with (i) dominant contribution by up-electrons and (ii) the susceptibility in the z-direction is about $8\times$ larger compared to the susceptibility in the $x$-direction.
The susceptibility anisotropy is $\Delta\chi_\mathrm{para}=253.68\,\mathrm{[10^{-6}, SI]}$, dominantly provided by the Fe atom, as the contribution of diamagnetic anisotropy is $\Delta\chi_\mathrm{vol,dia}=-1.04$\,[$10^{-6}$, SI] (Tab.~\ref{tab2}).

Expressing slope on atomic Fe concentration
\begin{align}
\frac{\chi_{xx}}{x_\mathrm{Fe}} &= 207.8 \,\, \mathrm{[10^{-6}, SI]}
\\
\frac{\chi_{zz}}{x_\mathrm{Fe}} &= 1729.9 \,\, \mathrm{[10^{-6}, SI]}.
\end{align}
It provides contribution of single Fe atom to the AMS 
\begin{equation}
\frac{\Delta \chi}{x_\mathrm{Fe}} = \frac{\chi_{zz}-\chi_{xx}}{x_\mathrm{Fe}}
= 1522 \,\, \mathrm{[10^{-6}, SI]}.
\end{equation}
in close agreement with the experimental slope 1512\,\,$\mathrm{[10^{-6}, SI]}$ (Eq.~(\ref{eq:conv:Deltachi_xFe})). 
In the calculation, the anisotropy originates from a very different strength of the orbital magnetic response of the Fe atom to magnetic fields along and perpendicular to the $c$-axis, similar to the diamagnetic AMS of undoped calcite. Since each supercell contains a single Fe atom, this anisotropy is a property of an individual Fe ion and does not require exchange between Fe atoms. This is consistent with the experiments: $\Delta k$ scales linearly with the Fe content \cite{Schmidt2006}, and its temperature dependence is the same for Fe contents ranging from about 900 to 11\,000\,ppm \cite{Schmidt2007}. On the other hand, our calculation with Fe atoms 4.94\,\AA\ apart and parallel Fe spins cannot by itself exclude a contribution of Fe--Fe interactions at higher Fe contents. The calculation demonstrates the importance of including the orbital magnetisation of Fe atoms for a proper description of the paramagnetic anisotropy.

\subsection{Temperature dependence of the paramagnetic anisotropy}
\label{sec:Tdep}
The calculated orbital anisotropy is the response of the electronic ground state ($T=0$) and does not depend on temperature. In contrast, \cite{Schmidt2007} found that the paramagnetic $\Delta k$ of Fe-bearing calcite increases by a factor $p_{77}=\Delta k(77\,\mathrm{K})/\Delta k(\mathrm{RT})=13.3$ upon cooling from room temperature to 77\,K. This is much more than the factor $300/77=3.9$ expected for a Curie-type $1/T$ dependence, and close to the factor $(300/77)^2=15.2$ expected for a $1/T^2$ dependence.

A $1/T^2$ dependence is the signature of the single-ion anisotropy of a paramagnetic ion \cite{AbragamBleaney1970}. The crystal field of the trigonally distorted octahedral site, together with spin-orbit coupling, splits the five ${S=2}$ states of Fe$^{2+}$ already in zero magnetic field. In the simplest (axial) form, this zero-field splitting is described by the spin Hamiltonian $H=D S_z^2$, with $z\parallel c$. When the thermal energy is large compared with the splitting, $k_BT\gg|D|$, the leading term of the susceptibility anisotropy is
\begin{equation}
\Delta\chi = \chi_{\parallel}-\chi_{\perp} \approx -\frac{C\,D\,(2S-1)(2S+3)}{10\,k_B T^2},
\label{eq:zfs}
\end{equation}
where $C$ is the Curie constant ($\chi=C/T$, Eq.~(\ref{eq:Curie})). The anisotropy is thus one power of $1/T$ steeper than the mean susceptibility, because it arises from the small difference in the thermal population of the split levels. Fitting only $D$ to the room-temperature ratio of anisotropy and mean paramagnetic susceptibility of Fe in calcite (1512/3404, Eq.~(\ref{eq:Deltachi_xFe}) and Sec.~\ref{sec:spinpara}), using the exact susceptibility of the spin Hamiltonian, gives ${D\approx-60}$\,K (about $-42$\,cm$^{-1}$; ${D<0}$ makes the $c$-axis the easy axis). This single parameter predicts ${p_{77}=12.4}$, close to the measured 13.3. The value of $D$ should be taken as effective, because the orbital moment of Fe$^{2+}$ in a trigonally distorted octahedron is only partly quenched. In a plot of $1/\chi$ versus $T$, the zero-field splitting appears as different paramagnetic Curie temperatures $\theta$, defined by the Curie--Weiss law ${\chi=C/(T-\theta)}$, for the two directions; at high temperature $\theta_\parallel=-D(2S-1)(2S+3)/15$ and $\theta_\perp=-\theta_\parallel/2$, whereas isotropic exchange between Fe ions shifts both by the same amount. \cite{Rochette1988} reported for two Fe-bearing calcites ${\theta_\parallel>0>\theta_\perp}$, in agreement with this sign pattern; for the sample in his Fig.~1b, ${\theta_\parallel\approx30}$\,K and ${\theta_\perp\approx-51}$\,K. Fitting the susceptibility of the spin Hamiltonian between 77 and 300\,K in the same way, i.e.\ by straight lines in $1/\chi$ versus $T$, reproduces both values with ${D\approx-37}$\,K, including their negative mean value $(\theta_\parallel+2\theta_\perp)/3\approx-24$\,K (the average over the three principal directions). This mean shift is a consequence of the curvature of $1/\chi(T)$ caused by the zero-field splitting and does not require interactions between Fe ions. The value ${D\approx-37}$\,K is smaller than the $-60$\,K obtained from the data of \cite{Schmidt2006, Schmidt2007}, which may reflect differences between the samples, e.g.\ in their Mn content, since Mn$^{2+}$ adds an almost isotropic paramagnetic contribution. For $D$ between $-37$ and $-60$\,K, the model gives $p_{77}$ between 14.5 and 12.4, which brackets the measured 13.3.

If a fraction $f$ of the room-temperature anisotropy were temperature independent, as in our calculation, the ratio would be $p_{77}\approx15.2(1-f)+f$. The measured ${p_{77}=13.3}$ thus implies ${f\lesssim0.15}$. Hence, the close agreement between the calculated orbital anisotropy and the experimental room-temperature slope is likely partly fortuitous, and the paramagnetic AMS at room temperature is dominated by the thermal population of the single-ion levels of Fe$^{2+}$. Both effects have the same origin, i.e.\ the anisotropic orbital response of the Fe ion in the trigonal crystal field, which spin-orbit coupling transfers to the spin. A quantitative description requires the calculation of the temperature-dependent population of these levels, e.g.\ by combining DFT with a crystal-field or spin-Hamiltonian model, which is beyond the scope of this work.

\section{Discussion}
\label{sec:discussion}

We demonstrate the diamagnetic susceptibility of carbonates can be obtained reliably within our DFT-based approach combined with perturbation theory. For calcite, the simulated susceptibilities and their anisotropy agree with available experimental susceptibility \cite{Schmidt2006}, indicating that (i) the underlying electronic structure is captured sufficiently well and (ii) the numerical settings required for converged diamagnetic response are tractable within standard workflows. Based on this validation, we provide improved reference values for magnesite and dolomite, for which diamagnetic susceptibilities are not straightforward to extract from experiments due to the ubiquitous presence of magnetic impurities (\cite{Schmidt2007};\cite{Biedermann2018}).

In contrast, the paramagnetic contribution from transition-metal dopants is substantially more demanding. It requires large supercells to represent dilute concentrations and to minimize spurious dopant-dopant interactions. While the computed spin moments of Fe and Mn dopants are robust and compare well with experiment, the paramagnetic anisotropy is far less straightforward. In our calculations, the anisotropy is controlled primarily by the orbital contribution, which is (i) difficult to isolate experimentally and (ii) difficult to converge numerically, as it tends to require very dense $k$-point sampling and exhibits slow convergence with respect to computational parameters. Moreover, the strong temperature dependence of the paramagnetic anisotropy observed experimentally \cite{Schmidt2007} shows that, at room temperature, it is dominated by the thermal population of the single-ion levels of Fe$^{2+}$ (Sec.~\ref{sec:Tdep}), which a ground-state calculation does not capture. 

Our results suggest that the dominant source of the diamagnetic anisotropy is not spin-orbit coupling, but an orbital mechanism tied to the crystallographic $c$-axis (i.e., an easier orbital response for electronic motion perpendicular to the $c$-axis). For the paramagnetic anisotropy, our preliminary calculations and the temperature dependence of the experimental data point to the anisotropy of individual Fe$^{2+}$ ions, whose orbital response is also tied to the $c$-axis, rather than to exchange between dopant atoms \cite{Schmidt2007}; at finite temperature, spin-orbit coupling transfers this orbital anisotropy to the spin through the zero-field splitting. A quantitative, predictive description therefore hinges on more accurate and more efficient access to orbital magnetisation and its field derivative, combined with the thermal population of the Fe$^{2+}$ levels. Achieving this will likely require methodological development beyond the present implementation (e.g., improved perturbative formalisms and/or computational techniques that reduce the cost of dense Brillouin-zone sampling) \cite{Gao2015}.

More broadly, magnetic doping of wide-band-gap host materials provides a versatile platform for solid-state magnetism, often producing magnetic responses that are not present in the undoped crystals. In this context, Fe-doped calcite is particularly interesting as a potential material system for quantum-technology defect engineering, motivating future work that combines improved orbital magnetization theory with more realistic (lower-concentration) defect models and a systematic study of temperature effects.

\section{Conclusions}

We calculated the diamagnetic susceptibility and AMS of calcite-group carbonates using DFT combined with perturbation theory, and we evaluated AMS contributions from both intrinsic diamagnetism and paramagnetic dopants. For calcite, the computed diamagnetic susceptibility and AMS agree with published single-crystal data, with values bracketed by the PBE and mBJ exchange-correlation potentials, showing that the diamagnetic AMS is captured reliably. For magnesite and dolomite, the raw experimental values are less negative than calculated, because the measured samples contain paramagnetic Fe and Mn. After subtracting the Curie contribution of the measured Fe and Mn contents, the experimental values agree with the calculated ones within about $1\times10^{-6}$\,[SI]. Hence, our calculations provide intrinsic diamagnetic reference values that can be combined with independent constraints on impurity content.

We further modelled Fe and Mn substitution in calcite using supercells. The computed moments reproduce the expected high-spin configurations (Fe$^{2+}$, 3d$^6$, $S=2$; Mn$^{2+}$, 3d$^5$, $S=5/2$), yielding effective moments $\mu_\mathrm{eff,Fe}=4.90\,\mu_B$ and $\mu_\mathrm{eff,Mn}=5.92\,\mu_B$. When the Fe and Mn contents are combined into a Fe-equivalent concentration, the calculated susceptibility agrees with the experimental data of calcite, dolomite and magnesite at low dopant concentrations. For Fe-doped calcite, the preliminary calculated paramagnetic susceptibility anisotropy per Fe concentration is comparable to the experimental slope, indicating that the observed AMS enhancement is linked to the anisotropy of the orbital motion of electrons of individual Fe ions.

Finally, we find that the diamagnetic AMS is orbital and linked to the crystallographic $c$-axis, and is not caused by spin-orbit coupling. The strong temperature dependence of the paramagnetic AMS is consistent with the single-ion anisotropy of Fe$^{2+}$ rather than with Fe--Fe exchange at dilute concentrations. However, improvement of the quantitative description of paramagnetic AMS is required by more efficient calculations of orbital magnetisation at realistic (dilute) dopant concentrations, combined with the thermal population of the Fe$^{2+}$ levels.

\section*{Conflict of Interest declaration}
The authors declare there are no conflicts of interest for this manuscript.

\acknowledgments
We acknowledge financial support by The Czech Science Foundation GACR (GA22-12828S), \textit{"New perspectives in magnetic fabric interpretation through 3D microstructural analysis, numerical modelling, and quantum mechanical description"}. J.H.\ acknowledges financial support by the FerrMion project of the Czech Ministry of Education, co-funded by the EU,
Project No. CZ.02.01.01/00/22\_008/0004591. 
Computational resources were provided by the ELIXIR-CZ project (ID:90255), part of the international ELIXIR infrastructure.


\section*{Data Availability Statement}
The analysis and plotting code, the key Wien2k ab-initio input and output files for all calculations presented in this study, and the collected diamagnetic and paramagnetic susceptibility values (both calculated and experimental) are available at~Zenodo (version 1.1), \mbox{\url{https://doi.org/10.5281/zenodo.24660686}} \cite{Hamrle2026}.

%
\bibliography{carbonate}
%

\appendix

\section{Concentration unit conversion}
\label{app:conc}

The Fe concentration in CaCO$_3$ in \cite{Schmidt2006, Schmidt2007} is expressed as a mass fraction 
\begin{equation}
w_\mathrm{Fe}=\frac{m_\mathrm{Fe}}{m_\mathrm{CaCO_3}}, 
\end{equation}
with $m_\mathrm{Fe}$ and $m_\mathrm{CaCO_33}$ being mass of Fe and CaCO$_3$ in the sample. 

Another expression of Fe concentration is atomic concentration $x_\mathrm{Fe}$ (also known as atomic fraction or atomic number density)
\begin{equation}
x_ {\mathrm{Fe}} = \frac{n_{\mathrm{Fe}}}{n_{\mathrm{CaCO3}}} 
=
\frac{m_\mathrm{Fe}\,M_\mathrm{CaCO_3}}{m_\mathrm{CaCO_3}\,M_\mathrm{Fe}}
=
w_\mathrm{Fe}\frac{M_\mathrm{CaCO3}}{M_\mathrm{Fe}}
\label{eq:Fefraction_w}
\end{equation}
where the number of moles density is
$n_{\text{Fe}} = \frac{m_{\text{Fe}}}{M_{\text{Fe}}}$
and
$n_{\text{CaCO3}} = \frac{m_{\text{CaCO3}}}{M_{\text{CaCO3}}}$
where the molar mass of Fe $M_{\text{Fe}} = 0.055845 \,\text{kg/mol}$ and the molar mass of formula unit of calcite (CaCO3)
$M_{\text{CaCO}_3} = 0.100087 \,\text{kg/mol}$.

The \cite{Schmidt2006, Schmidt2007} expresses of Fe concentration $c_\mathrm{Fe}$ in weight ratio in ppm, for example, 
$w_{Fe}$ = 1 ppm Fe = 1 $\mu$g Fe per gram of calcite = mass fraction of $10^{-6}$ of Fe inside CaCO$_3$. 
The same conversion is used for Mn ($M_\mathrm{Mn}=0.054938$\,kg/mol) and for the other hosts, where $M_\mathrm{CaCO_3}$ is replaced by the molar mass per cation site, i.e.\ $0.092200$\,kg/mol for dolomite (Ca$_{0.5}$Mg$_{0.5}$CO$_3$) and $0.084314$\,kg/mol for magnesite (MgCO$_3$). Thus, $x$ is always the atomic concentration per cation site.



\section{Magnetic susceptibility unit conversion}

The volume magentic susceptibility (dimensionless) is written
\begin{equation}
\chi^{(\mathrm{SI})}_\mathrm{vol}=\frac{M}{H_\mathrm{ext}} = \frac{\mu_0 M}{B_\mathrm{ext}} = \frac{\braket{B_\mathrm{ind}}}{B_\mathrm{ext}}
\end{equation}
assuming the magnetization $\mu_0 M=\braket{B_\mathrm{ind}}$ being much smaller compared to the external magnetic field, $\mu_0 M\ll B_\mathrm{ext}$.

Mass susceptibility writes:
\begin{equation}
\chi_{\rm mass}^{(\mathrm{SI})}=
\frac{1}{\rho}\chi^{(\mathrm{SI})}_{\rm vol} 
\label{eq:conv_ximass}
\end{equation}
where $\rho$ is the density of the material. The mass susceptibility is used in \cite{Schmidt2006, Schmidt2007}, with units $10^{-6}$ m$^3$/kg, SI.

Molar susceptibility (in units kg/mol) is
\begin{equation}
\chi_\mathrm{mol}^{(\mathrm{SI})}=
\chi_\mathrm{vol}^{(\mathrm{SI})} \frac{M}{\rho} 
= \chi_\mathrm{mass}^{(\mathrm{SI})} M
\label{eq:conv_ximol}
\end{equation}
where $\rho$ is mass density and $M$ is molar mass in units kg/mol.


Furthermore, Wien2k in NMR package calculates susceptibility in cgs units. To convert to SI, the factor $4\pi$ must be used:
\begin{align}
\chi^{(SI)}_\mathrm{vol} &= 4\pi \chi^{(cgs)}_\mathrm{vol}
\\
\chi^{(SI)}_\mathrm{mol} &= 4\pi \chi^{(cgs)}_\mathrm{mol}
\\
\chi^{(SI)}_\mathrm{mass} &= 4\pi \chi^{(cgs)}_\mathrm{mass}
\label{eq:}
\end{align}

Also, Wien2k expresses molar susceptibility $\chi^{(\mathrm{cgs})}_\mathrm{mol,W2k}$ in cgs units in 
 [10$^{-6}$ cm$^3$ mol$^{-1}$ cells $^{-1}$], which converts to SI volume susceptibility as 
 \begin{equation}
 \chi_\mathrm{vol}^{(\mathrm{SI})} [10^{-6}, \mathrm{SI}]
 =  
 \chi^{(\mathrm{cgs})}_\mathrm{mol,W2k} 
  [10^{-6} \mathrm{cm}^3 \mathrm{mol}^{-1} \mathrm{unitcell}^{-1}] \frac{4\pi}{V_\mathrm{mol}}
  \label{eq:conv_chivolw2k}
 \end{equation}
where $V_\mathrm{mol}$ is the molar volume (volume of one mol of unit cells, $V_\mathrm{mol} = V_\mathrm{unitcell} N_A$, $V_\mathrm{unitcell}$ is the volume of a unit cell) in units (cm$^3$/ unit cell) $N_A$ is Avogadro number and $4\pi$ is the conversion from cgs susceptibility to SI susceptibility.

\section{Curie-Weiss model}
\label{app:curie}

The volume susceptibility of paramagnetic impurities expressed by the Curie-Weiss model is
\begin{equation}
\chi_\mathrm{vol}^{\text{(SI)}}= \frac{\mu_0 N_A n_\mathrm{Fe} (\mu_{\rm eff}\mu_B)^2}{3 k_B T}
\label{eq:Curie-Weiss-appendix}
\end{equation}
where $n_\mathrm{Fe}$ is the mole density of Fe atoms 
(i.e.\ $N_A n_\mathrm{Fe}$ is the number of Fe atoms per unit volume), $N_A$ is Avogadro number, $ \mu_0$  
is vacuum permeability, $\mu_B$ is Bohr magneton and $k_B$ is Boltzmann constant. 

The effective magnetic moment
$\mu_\mathrm{eff}$ of the magnetic atom (e.g.\ Fe), in units $\mu_B$ is expressed as
\begin{equation}
\mu_\mathrm{eff} = 
\frac{1}{\mu_B} \sqrt{\frac{3k_BT \chi_\mathrm{vol}^{\mathrm{(SI)}}}{\mu_0 N_A n_\mathrm{Fe} }} 
\end{equation}
where $\chi^{\mathrm{(SI)}}_\mathrm{vol}=\chi^{\mathrm{(SI)}}_\mathrm{mass}\rho$ is the volume magnetic susceptibility and molar density of Fe ions $n_\mathrm{Fe}$ relates with weight concentration of Fe ions $w_\mathrm{Fe}$, Eq.~(\ref{eq:Fefraction_w})
\begin{equation}
n_{\mathrm{Fe}}
= x_\mathrm{Fe} n_\mathrm{CaCO_3} 
= x_\mathrm{Fe} \frac{\rho}{M_\mathrm{CaCO_3}}
= w_\mathrm{Fe} \frac{\rho}{M_\mathrm{Fe}}
\end{equation}
where $\rho$ is the density (mass density) of calcite.

For several kinds of paramagnetic ions $i$ (here Fe$^{2+}$ and Mn$^{2+}$), the Curie contributions add,
\begin{equation}
\chi_\mathrm{vol}^{\text{(SI)}}= \frac{\mu_0 \mu_B^2}{3 k_B T\,V_\mathrm{fu}} \sum_i x_i\, \mu^2_{\mathrm{eff},i}
= \frac{\mu_0 \mu_B^2\, \mu^2_\mathrm{eff,Fe}}{3 k_B T\,V_\mathrm{fu}}\, x_\mathrm{eff},
\label{eq:Curie-sum}
\end{equation}
where $V_\mathrm{fu}$ is the volume per formula unit (per cation site) and $x_\mathrm{eff}$ is the Fe-equivalent concentration, Eq.~(\ref{eq:xeff}). With $V_\mathrm{fu}$ of calcite, dolomite and magnesite, Eq.~(\ref{eq:Curie-sum}) gives at $T=300$\,K the slopes $\chi/x_\mathrm{eff}=3404$, 3922 and $4477\times10^{-6}$\,[SI], respectively. The experimental mass susceptibilities were converted to volume susceptibilities using the densities 2710, 2860 and 3000\,kg/m$^3$ for calcite, dolomite and magnesite \cite{Schmidt2006, Schmidt2007}.

It provides the final relation for the effective magnetic moment
\begin{equation}
\mu_\mathrm{eff} = 
\frac{1}{\mu_B} \sqrt{\frac{3k_BT \chi^{\mathrm{(SI)}}_\mathrm{mass}M_\mathrm{Fe} }{\mu_0 N_A w_\mathrm{Fe}}} 
\label{eq:conv:mueff_w}
\end{equation}
with the experimental slope for Fe doping in CaCO$_3$ \cite{Schmidt2006} "the increase of susceptibility is $2.3 \times 10^{-10}$\,m$^3$/kg per 100 ppm Fe"
meaning $\chi_\mathrm{mass}^\mathrm{(SI)}=2.3\times 10^{-10}$\,m$^3$/kg and related weight concentration of Fe ions $w_\mathrm{Fe}=100\, \mathrm{[ppm]} = 10^{-4}$, providing effective magnetic moment of single Fe ion in CaCO$_3$ matrix to be $\mu_\mathrm{eff}=4.94$\,$\mu_B$.


\section{Anisotropy of magnetic susceptibility unit conversion}

In \cite{Schmidt2006}, Sec.~4.2 and Fig.~8, the scaling of magnetic anisotropy with Fe mass concentration is experimentally determined 
\begin{equation}
\Delta\chi^{\mathrm{(SI)}}_\mathrm{mass} =k_{1}-k_3=\Delta k [\mathrm{m^3/kg}] = w_\mathrm{Fe} \mathrm{[ppm]} \times (1\pm 0.1)\times 10^{-12}\,\mathrm{[m^3/kg/ppm]} 
\end{equation}
We convert this expression of magnetic anisotropy into a ratio of volume susceptibility and atomic concentration $x_\mathrm{Fe}$ 
\begin{equation}
\frac{\Delta \chi^{\mathrm{(SI)}}_\mathrm{vol}}{x_\mathrm{Fe}} 
= 
\frac{\Delta\chi^{\mathrm{(SI)}}_\mathrm{mass}\,\rho}{w_\mathrm{Fe} M_\mathrm{CaCO_3}/M_\mathrm{Fe}} = 1512\,\, [10^{-6}, \mathrm{SI}],
\label{eq:conv:Deltachi_xFe}
\end{equation}
which expresses the contribution of a single Fe ion to the anisotropy of magnetic susceptibility.
In this conversion, we have substituted $\Delta\chi^{\mathrm{(SI)}}_\mathrm{mass} /w_\mathrm{Fe} = 1.0\times 10^{-12} \times 10^6$[kg/m$^3$] (value of $\Delta k/w_\mathrm{Fe}$) and 
where $\Delta \chi^{\mathrm{(SI)}}_\mathrm{vol} = \chi^{\mathrm{(SI)}}_{\mathrm{vol},zz} - \chi^{\mathrm{(SI)}}_{\mathrm{vol},xx}$, which equals $k_1-k_3$ because, for Fe-bearing calcite, the principal axis of $k_1$ is parallel to the $c$-axis.

\section{Comparison of individual samples with the calculation}
\label{app:residuals}
\begin{figure}
\centering
\includegraphics[width=0.85\linewidth]{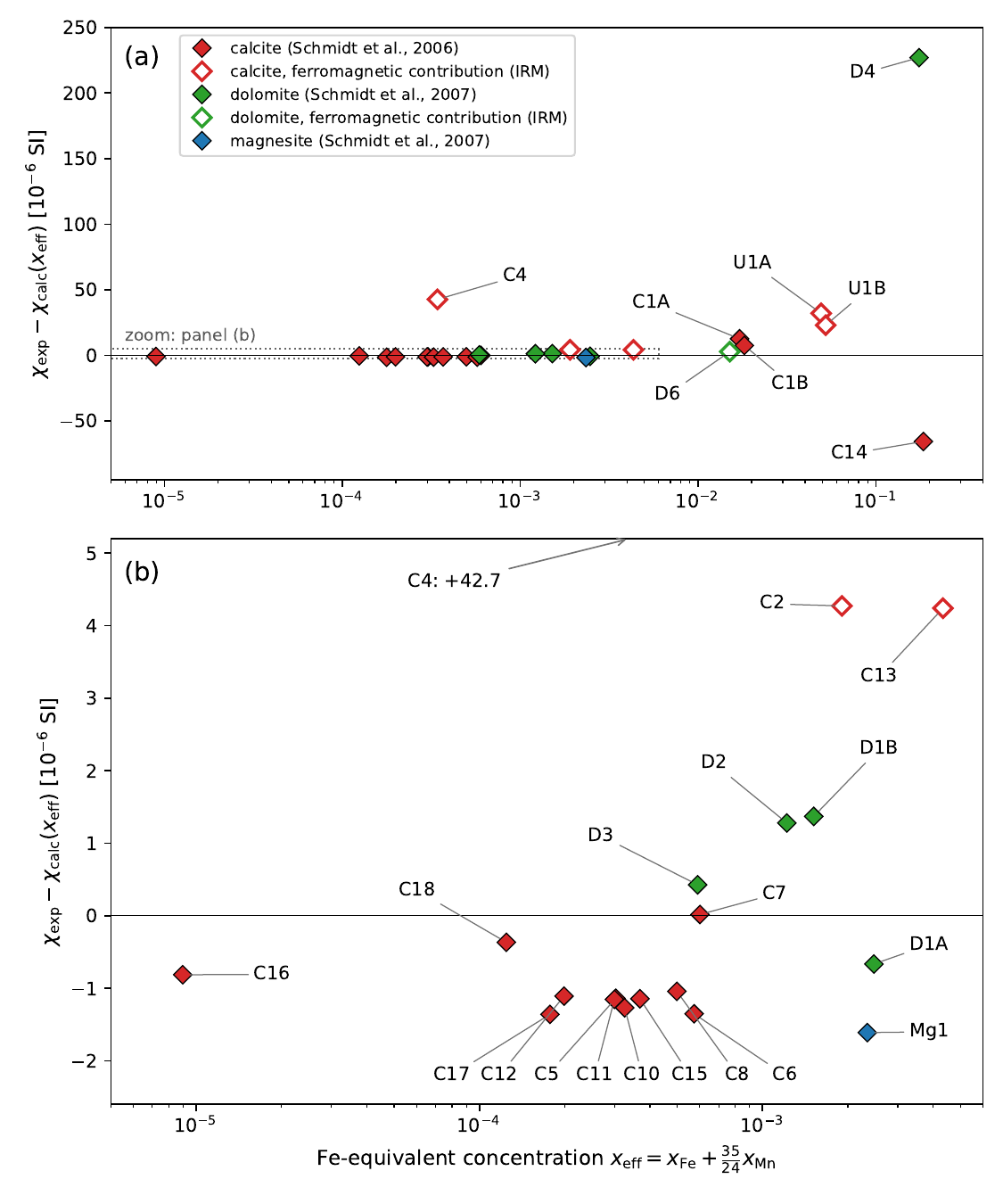}
\caption{Difference between measured \cite{Schmidt2006, Schmidt2007} and calculated susceptibility, $\chi_\mathrm{exp}-\chi_\mathrm{calc}(x_\mathrm{eff})$, versus the Fe-equivalent concentration $x_\mathrm{eff}$ for calcite (red), dolomite (green) and magnesite (blue). Open diamonds: samples with a ferromagnetic contribution detected by IRM. (a) All samples; the dotted rectangle marks the range shown in (b). (b) Zoom into the low-concentration range.}
\label{fig:residuals}
\end{figure}

Fig.~\ref{fig:residuals} shows, for each sample of \cite{Schmidt2006, Schmidt2007}, the difference between the measured susceptibility and the calculated one at the same Fe-equivalent concentration $x_\mathrm{eff}$ (PBE diamagnetic susceptibility of the host plus the Curie contribution, Eq.~(\ref{eq:Curie-sum})). Calcite samples without a ferromagnetic contribution lie about $-1\times10^{-6}$\,[SI] below zero, which is the difference between the PBE value ($-11.06\times10^{-6}$\,[SI]) and the intrinsic diamagnetic susceptibility of calcite ($-12.09\times10^{-6}$\,[SI], \cite{Schmidt2006}). Samples with a ferromagnetic contribution detected by IRM lie above zero. Large deviations occur only at high concentrations: the Mn-rich calcite C14 and the Fe-rich dolomite D4.

\end{document}